\documentstyle[12pt]{article}
\title{Examples of $q$-regularization.\\}

\author{Suemi Rodr\'\i guez-Romo\\Centre of Theoretical Research\\
Facultad de Estudios Superiores Cuautitl\'an, UNAM \\
Apdo. Postal 95, Unidad Militar, Cuautitl\'an Izcalli\\
Estado de M\'exico, 54768 M\'exico.$^{\ast}$.}

\date{}

\begin{document}

\maketitle

\renewcommand{\thefootnote}{\fnsymbol{footnote}}
\setcounter{footnote}{-1}
\footnote{$\hspace*{-6mm}^{\ast}$
e-mail: suemi@fis.cinvestav.mx\\
$\hspace*{1.2cm}$ suemi@servidor.dgsca.unam.mx}
\renewcommand{\thefootnote}{\arabic{footnote}}
{\small {\bf Abstract. An introduction to Hopf algebras as a tool for the
regularization of relevant quantities in quantum field theory is given. We
deform algebraic spaces by introducing $q$ as a regulator of a
non-commutative and non-cocommutative Hopf algebra. Relevant quantities are
finite provided $q\neq 1$ and diverge in the limit $q\rightarrow 1$. We
discuss $q$-regularization on different $q$-deformed spaces for
$\lambda\phi^4$ theory as example to illustrate the idea} .\\

}

\newpage

\section{Introduction}
Thanks to the work done in expressing vector bundles, forms, integration,
etc., on locally compact topological spaces $X$, entirely in terms of the
algebra $C(X)$ of complex continuous functions on $X$ vanishing at infinity
which forms a commutative $C^{\ast}$ algebra; a generalization of ordinary
geometry can be introduced. Namely, when expressed in terms of a $C^{\ast}$
algebra the above cited notions make sense even when the $C^{\ast}$ algebra
is not commutative, therefore not of the form $C(X)$ \cite{Co}. The simplest
non-commutative geometries that have been studied are non-commutative and
non-cocommutative Hopf algebras, corresponding to both quantization and
curvature.\

Meanwhile in classical mechanics states are points of a manifold $M$ and
observables are functions on $M$; in the quantum case, states are
one-dimensional subspaces of a Hilbert space $H$ and observables are operators
in
$H$. Observables, in both classical and quantum mechanics, form an
associative algebra, which is commutative in the classical case and
non-commutative in the quantum case. So, we can think of quantization as a
procedure that replaces the classical algebra of observables by a
non-commutative quantum algebra of observables. The non-commutative Heisenberg
algebra, i.e. the algebra that comes up because  momentum and space are not
simultaneously measurable (so called Heisenberg's uncertainty principle), is
the best example to illustrate this idea. Generally speaking,  it is expected
that  even using non commutative geometry, one might nevertheless  extend our
regular notions of symmetry to the quantum world. If we consider the space of
states endowed by a group structure, the functions on this are observables. To
quantize such a system one has to construct a non-commutative associative
algebra of functions on a locally compact topological group space; i.e. a
quantum group \cite{Dr}.\

Thinking about quantization of the space-time metric itself, where we cannot
use path integration techniques to express quantization in terms of classical
fields; we claim the assumption of a smooth manifold structure for space-time
to be meaningless in extremely small scales from the experimental viewpoint.
The problem is that the finer the accuracy in the observation we ask for, the
heavier the test particle we need; eventually the space-time curvature due to
both the test particle and the space-time itself can be of the same magnitud.
In this context, by relaxing the assumption of smoothness of the space-time
manifold and introducing non-commutative algebraic geometry, we propose
a scheme called $q$-regularization, so we can regulate relevant quantities
in field theory before renormalising. $q$ (being $q^2\neq -1$) parametrizes the
deformation to the non-commutative and
non-cocommutative framework in which relevant quantities in quantum field
theories
are finite for $q\neq 1$, and reduce to the unregulated, divergent, physical
theory as $q\rightarrow 1$. Namely, as well as in dimensional regularization
we interpolate consistently to dimension $4-\epsilon $ where the relevant
quantities are finite  (these would be infinite at dimension four); in
$q$-regularization we extend relevant quantities in quantum  field theory to a
non-commutative and non-cocommutaive Hopf algebra or  quantum group (by
introducing the parameter $q$ ) where the relevant quantities are finite
(these would be infinite at $q=1$; i.e. in $C(X)$, the commutative limit).\

We present two examples, the first one is constructed in a four dimensional
representation of a particular non-commutative space previously
reported \cite{Mj2}. The second example is proposed having in mind
$q$-spinors (two dimensional objects with the generators of $A^{2/0}_{q}$,
Manin's quantum plane \cite{Ma}, as entries) constructed by the projective
representation of the Heisenberg algebra, they are braided in a very specific
way to obtain a $q$-deformed space.\

Second example is intended as a first step to approach $q$-regularization in
$q$-Minkowski space-time. We work out this example in a $q$-deformed space
which can be related with both, first example's $q$-mutator algebra and
previously reported \cite{Ca1}\cite{Ca2} braided two copies of Manin's
quantum planes. Since we do not impose reality conditions, among others, we
are not working in anyway in $q$-Minkowski space-time.\

For the second example we want to learn more about the symmetries of our
measure, we study a projection in the $q$-deformed space used and its
relation to the $SU_{q}(2)$ measure. Moreover, we analyze the null directions
of the corresponding Hopf algebra that lead to a $q$-deformed Galilei group.\

This paper is organized as follows; in Section 2 we construct the Manin
quantum plane out of the non-commutative Heisenberg algebra and introduce
the $q$-spinors as a way to link $q$-regularization scheme with physically
meaningful concepts. In section 3, we present two examples of
$q$-regularization on $q$-deformed Euclidean spaces for $\lambda\phi^4$
theory. Our scheme can only be carried out in a very particular basis for
functions defined on the $q$-deformed spaces chosen such that we end with a
Haar weight that reduces to an ordinary integration. Further work should be
done to generalize this. Finally, in order to learn about desired properties
of symmetry in this $q$-regularization we study the zero time projection of
the measure we have just introduced in second example in terms of the
$SU_{q}(2)$ measure and the null directions of the Hopf algebra that lead to
a $q$-deformed Galilei group. The quantum Galilei group has been found as
symmetry in condensed matter \cite{Bo}.\

\section{From Heisenberg algebra to $q$-spinors.}

The goal of this section is to link non-commutative Heisenberg algebra with
two co-cycles and $q$-spinors as defined by Manin \cite{Ma}. Let us start by
the fundamental Heisenberg commutator algebra generated by translations on
phase space $({\bf r},{\bf p})$;
\begin{eqnarray}
&&\left[r^{i},p^{j}\right] = i \hbar \delta^{ij} \label{line1} \\
&&\left[r^{i},r^{j}\right] = \left[p^{i},p^{j}\right]=0 .\nonumber
\end{eqnarray}
We propose the following translation operator on phase space;
\begin{equation}
U({\bf a} ,{\bf b} )=e^{i ({\bf a}\cdot {\bf p} -
{\bf b} \cdot {\bf r} )/ \hbar }
\mbox{ where ${\bf a }$ and ${\bf b }\in R^n$ .}
\end{equation}
In a ray or projective representation, eq(2) obeys the composition law
\cite{Dj}.\

{\large\begin{equation}
U({\bf a}_{2},{\bf b}_{2})\cdot U({\bf a}_{1},{\bf b}_{1})= e^{ \left[2\pi
i \alpha_{2}({\bf r} ;
({\bf a}_1 ,{\bf b}_1),({\bf a}_2 ,{\bf b}_2))\right]}\cdot U({\bf a}_1
+{\bf a}_2 ,{\bf b}_1 +{\bf b}_2 ),\\
\end{equation}}
where  ${\bf a}_{1},{\bf b}_{1},{\bf a}_{2},{\bf b}_{2}\in R^n$ and, for a
free particle in quantum
mechanics,  the two co-cycle
$\alpha_2$  for translations in the phase space is given by
\begin{equation}
2 \pi \alpha_{2}\left({\bf r} ;
({\bf a}_1 ,{\bf b}_1),({\bf a}_2 ,{\bf b}_2)\right) =
\frac{1}{2\hbar}({\bf a}_1 \cdot {\bf b}_2 -{\bf a}_2\cdot {\bf b}_1 ) .\\
\end{equation}
Let us now consider the following infinitesimal Galilei transformation
\cite{Dj}
\begin{eqnarray}
{\bf r'}={\bf r}+{\bf a}_1={\bf r} +\hbar {\bf u} & & {\bf r''} ={\bf r} +
{\bf a}_2 ={\bf r} \\
{\bf p'}={\bf p} + {\bf b}_1 ={\bf p} & & {\bf p''}={\bf p}+{\bf b}_2 =
{\bf p} +\hbar {\bf u}, \nonumber
\end{eqnarray}
\nonumber
 where {\bf u }is a unit vector in $R^{n}$.\

If we define
{\Large\begin{equation}
q=e^{-i\hbar} ,
\end{equation}}
impose eq(5)  as symmetry in eq(4) and substitute the result in eq(3),
it is straightforward to prove that
\begin{equation}
U(\hbar\bf u ,\rm0)U(0,\hbar{\bf u})=qU(0,\hbar {\bf u})U(\hbar \bf u,\rm 0)
\end{equation}
is a realization of $A_{q}^{(2/0)}$; i.e. this fulfills the non-commutative
algebra of the Manin's quantum plane \cite{Ma}.\

Like other authors \cite{Ca1}\cite{Ca2}, we call the following two dimensional
object a $q$-spinor (more properly Weyl $q$-spinor).
\begin{equation}
Z^{\rho}=
\left[
\begin{array}{c}
Z^1 \\
Z^2
\end{array}
\right]=
\left[
\begin{array}{c}
U(\hbar {\bf u} ,\rm 0)\\
U(0,\hbar {\bf u})
\end{array}
\right]
\mbox{, i.e. $\rho =1,2$ . }
\end{equation}
In example 2 we use an approach \cite{Ca1} \cite{Ca2} in which the
$q$-deformed space can be related with the tensor product representation of
two $q$-spinor spaces called $(Z^{i} ,\tilde{Z}^{i})$. A pair of $q$-spinors
$(i=1,2)$ is introduced in each space.
Hereafter greek indices are for spinor suffix and roman ones for different
spinors. Besides it is required the following braiding
\begin{equation}
Z^{i}\tilde{Z}^{j}= \hat{R}^{ij}_{j'i'}\tilde{Z}^{j'}Z^{i'}
\end{equation}
where $\hat{R}^{ij}_{j'i'}$ is the Yang-Baxter matrix for $ SL_{q}(2,C)$.

\section{Examples on $q$-regularization}

In this section we present two examples of $q$-regularization for
$\lambda\phi^4$
theory on two apparently different $q$-deformed spaces, both Euclidean. The
first case involves a four dimensional version of a Hopf algebra previously
reported \cite{Mj2}; we propose to extend momenta internal to Feynman loops
to a non-commutative structure. Second example involves a braided 4-dimensional
representation of Manin's quantum plane
(so called $q$-spinors) where some particular transformations on the
generators of this Hopf algebra relates to the one used in example 1.
Actually, example 1 is posed in order to better explain example 2 which is
considered as a preliminary step for formulating $q$-regularization on
$q$-Minkowski space-time.\

{\bf EXAMPLE 1}. From reference 3, let us consider the Hopf algebra ${\it L}$
generated by $\left(l_1,l_2,l_3,l_4\right)$ and
\begin{equation}
\left[l_k,l_j\right]=il_jQ', \mbox{ for k=2,4 and j=1,3}
\end{equation}
where $Q'=\sqrt{1-q}$. Define on this, the antipode map as
\begin {equation}
S\left(l_k\right) = -l_k  \;\;\;\;\;
S\left(l_j\right) = -q^{-i}l_jq^{-l_k/Q'},
\end{equation}
the coproduct map is given by
\begin{equation}
\bigtriangleup l_k = l_k\otimes 1+1\otimes l_k \;\;\;\;
\bigtriangleup l_j = l_j\otimes 1+q^{\frac{l_k}{Q'}}\otimes l_j
\mbox{ and}
\end{equation}
the counit
\begin{equation}
\epsilon\left(l_k\right)=\epsilon\left(l_j\right)=0.
\end{equation}
Additionally, ${\it L}$ can become a $C^{\ast}$ algebra if we define
\begin{equation}
l^{\ast}_k=l_k\;\;\;\;\;l^{\ast}_j=l^{\ast}_jq^{i/2}
\end{equation}
For every finite-dimensional Hopf algebra there is an invariant integration,
the Haar weight $\int$, unique up to normalization.\

A basis $B^{a_1,...a_4}=e^{ia_1l_1}...e^{ia_4l_4}$ where $a_n\in C$, being $C$
the complex is chosen, then the dual basis $D_{a'_1,...a'_4}$ is given via
\begin{equation}
B^{a_1...a_4}D_{a'_1...a'_4}=\delta(a'_1-a_1)...\delta(a'_4-a_4),\;\;\;
a'_n\in C
\end{equation}
where the Dirac delta functions $\delta$ have been defined with respect to the
usual Lebesgue integration, then it is straightforward, by analogy with the
case of finite dimensional Hopf algebras \cite{La}, to prove \cite{Su}
\begin{equation}
\int\int f=\left[2\pi\delta(0)\right]^{k}\int
\prod_{j}da'_jf'(0,a'_j(1-q^{-i}))
\end{equation}
for all $j$  and $f$ suitable of being written on the basis $B^{a_1...a_4}$.
In case $q\neq 1$ and assuming proper analycity and decay of $f'$
(the Fourier transform of the Wick ordered function $f$), eq(16) might be
finite for suitable $f$. If $q=1$ eq(16) certainly diverges.\

We propose, from $\lambda\phi^4$ theory, to $q$-regularize the vertex
corrections with
contributions given by
\begin{equation}
\Gamma(s)=\frac{\left(-i\lambda\right)^2}{2}\int\int\frac{d^4l}{(2\pi)^4}
\frac{i}{(l-p)^2-\mu^2_0-i\epsilon}\frac{i}{l^2-\mu^2_0+i\epsilon}
\end{equation}
where s is any of the Mandelstam variables. These corrections diverge
logarithmically.\

Let us extend the internal momentum in the Feynman loop in $\Gamma(s)$ to the
non-commutative algebraic framework by considering instead of the standard
Lebesgue integration, the Haar weight above defined on the basis
$B^{a_1...a_4}$, thus;
\begin{equation}
\Gamma_q(s)=
\end{equation}
$$
\frac{\lambda^2_0\delta(0)}{2(2\pi)^3}
\int\frac{d^jl'_j}
{\left[p^2_k+\left(l'_j(1-q^{-i})-p_j\right)^2-
\mu^2_0+i\epsilon\right]\left[\left(l'_j(1-q^{-i})\right)^2-\mu^2_0+
i\epsilon\right]}
$$
where $l'_j$ are the odd components of the dual internal momentum that was
extended to non-commutative geometry and $p_k(p_l)$ are the even (odd)
components of the external momentum in standard Euclidean commutative four
dimensional space-time. Unless $q=1$, eq(18) is finite, thus we have a
regularization scheme. An additional attempt of $q$-renormalization has
recently been presented \cite{Su}. Since we extend to the non-commutative
framework only the internal momenta degrees of freedom, the lack of locality,
consequence of this extention has not experimental consequences in this case.
\cite{Fr}.\

{\bf EXAMPLE 2}. Let us consider the Hopf algebra ${\it H}$ generated by 1 and
$(a,{\bar a},b,{\bar b})$ such that;
\begin{equation}
\left[b,{\bar b}\right]=0,\;\; \;\;\;
\left[a,{\bar a}\right]=2(q^{-1}-q)q^{\frac{1}{2Q'}(b+3{\bar b})}
\end{equation}
$$
\left[{\bar b},a\right]=\left[b,a\right]=2Q'{\bar a},\;\;\;\;\;
\left[{\bar b},{\bar a}\right]=\left[b,{\bar a}\right]=2Q'a .
$$
The coproduct map $\bigtriangleup $ in this Hopf algebra is
\begin{equation}
\bigtriangleup a=a\otimes 1+q^{\frac{b}{Q'}}\otimes a  \;\;\;\;\;
\bigtriangleup b=b\otimes 1+1\otimes b
\end{equation}
$$
\bigtriangleup {\bar a}={\bar a}\otimes 1+q^{\frac{{\bar b}}{Q'}}\otimes
{\bar a}  \;\;\;\;\;\bigtriangleup {\bar b}={\bar b}\otimes 1+1\otimes
{\bar b},
$$
the antipode map $S$ is
\begin{eqnarray}
S(a) &=& \frac{1}{2}
\left\{-(q^{-2}+q^2)aq^{-\frac{b}{Q'}}+(q^2-q^{-2}){\bar a}
q^{-\frac{b}{Q'}}\right\}  \\
S({\bar a}) &=& \frac{1}{2}
\left\{(q^2-q^{-2})aq^{-\frac{{\bar b}}{Q'}}-(q^{-2}+q^2){\bar a}
q^{-\frac{{\bar b}}{Q'}}\right\}\nonumber \\
S(b)=-b & & S({\bar b})=-{\bar b} \nonumber,
\end{eqnarray}
and finally the counit map $\epsilon $ is
\begin{equation}
\epsilon (a)=\epsilon({\bar a})=\epsilon(b)=\epsilon({\bar b})=0
\end{equation}
Furthermore, we can make this into a $\ast$-algebra via
$$
b^{\ast}=b,\;\;\;{\bar b}^{\ast}={\bar b}\;\;\;
a^{\ast}=aq^{i/2},\;\;\;{\bar a}^{\ast}={\bar a}q^{i/2}
$$
iff q is a primitive root of unity such that $q^4=1$.\

We would like to relate ${\it H}$ with
\begin{equation}
X^{ij}=\tilde{Z}^{i}Z^{j} \; \in A^{2/0}_{q}\otimes A^{2/0}_{q}\;\; i,j=1,2;
\end{equation}
where $\tilde{Z}^i$ and $Z^j$ where introduced in Section 2 (eq(8) and
eq(9)).
It is straightforward to prove that $A^{2/0}_q\otimes A^{2/0}_q$ is isomorphic
to the real algebra generated by 1 and (A,$\bar{A}$,B,$\bar{B}$), where
\begin{equation}
A=X+Y ,\;\;\;\bar{A}=X-Y ,\;\;\;B=Z+T ,\;\;\;\bar{B}=Z-T
\end{equation}
and
\begin{equation}
X=q^{-1/2}X_{11},\;\;\;Y=q^{-1/2}X_{12},\;\;\;Z=\frac{q^{-1}X_{21}-qX_{22}}
{\sqrt{q+q^{-1}}},\;\;\;T=\frac{X_{21}+X_{22}}{q\sqrt{q+q^{-1}}}.
\end{equation}
To relate $H$ with $A^{2/0}_q\otimes A^{2/0}_q$ let us rewrite the
$(A,\bar{A},B,\bar{B})$ generators, for $q\neq 1$, as follows\

{\large\begin{equation}
A=a ,\;\;\;\bar{A}=\bar{a} ,
\end{equation}
$$
 B=q^{\frac{b}{Q'}},\;\;\;
\bar{B}=q^{\frac{\bar{b}}{Q'}}.
$$}
On the other hand, it is straightforward to prove that in ${\it H}$,
$((a+\bar{a})$, $(a-\bar{a})$, $b$, $\bar{b})$ corresponds to the algebra
${\it L}$ with generators $(l_1,l_2,l_3,l_4)$ defined in example 1. If
$q\rightarrow 1$, the algebra becomes the commutative algebra of functions on
the space generated by ($a,\bar{a},b,\bar{b}$) and the unit.\

Like in example 1, we proceed defining the Haar measure $\int \int$ as a map
${\it H}\rightarrow C$, such that
\begin{equation}
\int \int f=f_{(1)}\int \int f_{(2)}\;\;\;\forall f\in {\it H}
\end{equation}
Here we have expressed the action of $\bigtriangleup$ on f as
$\bigtriangleup f=f_{(1)}\otimes f_{(2)}$. We remark that it is
well known in  the theory of Hopf algebras \cite{Mj1} that
eq(27) is the dual formulation of the usual left invariance.\

By analogy with the case of finite dimensional Hopf algebras \cite{La},
we use the following formal expression for eq(27)
\begin{equation}
\int \int f=Tr_{\it H}L_{f} S^2
\end{equation}
where $L_{f}$  stands for f acting by left multiplication on ${\it H}$.\

{}From eq(21) follows
\begin{equation}
S^2(a-{\bar a})=w^{-1}(a-{\bar a})\;\;\;;\;\;
S^2(a+{\bar a})=w^{-1}(a+{\bar a})\;\;\;;\;\;S^2b=b
\;\;\;\;;\;\;S^2\bar{b}=\bar{b}
\end{equation}
where $w^{-1}=f(q)$ and $lim_{q\rightarrow 1}w^{-1}=1$. This shall be used
below.\

To compute $\int \int$ we propose the following basis in ${\it H}$:\

{\large\begin{equation}
F^{\lambda_1 \lambda_2 , \lambda_3 \lambda_4 , \lambda_5 \lambda_6}=
(F^{\lambda_1 \lambda_2},F^{\lambda_3 \lambda_4},
F^{\lambda_5 \lambda_6})=
\end{equation}}
$$
\left(e^{i\lambda_1 \bar{b}}e^{i\lambda_2 \frac{(a-\bar{a})}{2}},
e^{i\lambda_3 \bar{b}}e^{i\lambda_4 \frac{(a+\bar{a})}{2}},
e^{i\lambda_5 \bar{b}}e^{i\lambda_6 b}\right)
$$
where
$$
F^{\lambda_1 \lambda_2 , \lambda_3 \lambda_4 , \lambda_5 \lambda_6}
\in {\it H}
\mbox{ and } (\lambda_1 ,\lambda_2 ,\lambda_3 ,\lambda_4 ,\lambda_5 ,\lambda_6)
\in R .
$$
We associate to
$F^{\lambda_1 \lambda_2 , \lambda_3 \lambda_4 , \lambda_5 \lambda_6}$
a dual basis
$F_{\lambda'_1 \lambda'_2 , \lambda'_3 \lambda'_4 , \lambda'_5 \lambda'_6}
\in (A^{2/0}_{q}\otimes A^{2/0}_{q})^{!}$
where $({A}^{2/0}_{q}\otimes A^{2/0}_{q})^{!}$ is the dual Hopf
algebra of ${A}^{2/0}_{q}\otimes A^{2/0}_{q}$, such that
\begin{equation}
F^{\lambda_1 \lambda_2 , \lambda_3 \lambda_4 , \lambda_5 \lambda_6}
F_{\lambda'_1 \lambda'_2 , \lambda'_3 \lambda'_4 , \lambda'_5 \lambda'_6}=
\left(
\delta (\lambda'_1-\lambda_1)\delta (\lambda'_2-\lambda_2),
\right.
\end{equation}
$$
\left.
\delta (\lambda'_3-\lambda_3)\delta (\lambda'_4-\lambda_4),
\delta (\lambda'_5-\lambda_5)\delta (\lambda'_6-\lambda_6)
\right)
$$
In basis eq(30) we have introduced six parameters $\lambda_i$, one for each
generator involved. They are dual variables to the non-commutative parameter.\

{\bf Theorem 1}. The Haar weight
$\int\int F^{\lambda_1 \lambda_2 , \lambda_3 \lambda_4 , \lambda_5 \lambda_6}$
defined in eq(28), for a basis
$F^{\lambda_1 \lambda_2 , \lambda_3 \lambda_4 , \lambda_5 \lambda_6}$
chosen as in eq(30), reduces to an ordinary integration.\

{\bf Proof}. From eq(19) we know that
$$
\left[\bar{b},(a-\bar{a})\right]=-2Q'(a-\bar{a})
$$
$$
\left[\bar{b},(a+\bar{a})\right]=2Q'(a+\bar{a})
$$
$$
\left[\bar{b},b\right]=0
$$
Note that $\frac{(a+\bar{a})}{2}=X$, $\frac{(a-\bar{a})}{2}=Y$ in eq(25).
Substituting the basis given by eq(30) in eq(28) and using the Glaube
formula for operators we obtain the ordinary integral
\begin{equation}
\int \int F^{\lambda_1 \lambda_2 ,\lambda_3 \lambda_4 ,\lambda_5 \lambda_6}=
\left(\int^{\infty}_{-\infty}d\lambda'_1d\lambda'_2\delta
(\lambda'_1-(\lambda_1+\lambda'_1))
\delta(\lambda'_2-(\lambda_2e^{-2i\lambda'_1Q'}+w^{-1}\lambda'_2)),
\right.
\end{equation}
$$
\int^{\infty}_{-\infty}d\lambda'_3d\lambda'_4\delta
(\lambda'_3-(\lambda_3+\lambda'_3))
\delta(\lambda'_4-(\lambda_4e^{2i\lambda'_3Q'}+w^{-1}\lambda'_4)),
$$
$$
\left.
\int^{\infty}_{-\infty}d\lambda'_5d\lambda'_6\delta
(\lambda'_5-(\lambda_5+\lambda'_5))
\delta (\lambda'_6-(\lambda_6+\lambda'_6))\right)
$$
Q.E.D.\

The basis $
F^{\lambda_1 \lambda_2 , \lambda_3 \lambda_4 , \lambda_5 \lambda_6}$
in eq(30) admits an expression in terms of the $q$-spinor defined in eq(8) out
of the projective representation for the Heisenberg algebra. Furthermore,
this basis can be rewritten in terms of $q$-Majorana spinors built using
$q$-Weyl spinors in analogy with the commutative algebraic formulation. As a
result of this we can show how does not matter if we think in terms of
integrating out non-commutative light cone coordinates, Weyl $q$-spinors or
Majorana $q$-spinors degrees of freedom; the result is exactly the same.
Furthermore, the Haar measure $\int \int$ defined on ${\it H}$ can be written
in terms of ordinary integration.\

{\bf Theorem 2}. For a suitable $f\in {\it H}$
 that can be expressed on the basis
$F^{\lambda_1 \lambda_2 , \lambda_3 \lambda_4 , \lambda_5 \lambda_6}$
given in eq(30) (or any of their different $q$-spinor representations),
$\int\int f$ as defined in eq(28) contains a component that can be
$q$-regularized, i.e. is finite provided $q\neq 1$, but infinite in the
limit $q=1$.\

{\bf Proof.}\

It is straightforward to show that for any function f
defined on ${\it H}$ with basis
$F^{\lambda_1 \lambda_2 , \lambda_3 \lambda_4 , \lambda_5 \lambda_6}$
the following transformation holds
\begin{equation}
f=:f':=\int^\infty_{-\infty}d\lambda_1d\lambda_2\tilde{f}
(\lambda_1,\lambda_2)F^{\lambda_1 \lambda_2}+
\end{equation}
$$
\int^\infty_{-\infty}d\lambda_3d\lambda_4\tilde{f}
(\lambda_3,\lambda_4)F^{\lambda_3 \lambda_4}+
\int^\infty_{-\infty}d\lambda_5d\lambda_6\tilde{f}
(\lambda_5,\lambda_6)F^{\lambda_5 \lambda_6}
$$\

where we express f as a normal ordered form of f',
in terms of the generators. Namely putting $\bar{b}$ to the left of $a$,
$\bar{a}$ and b in the light cone coordinate approach;
$\{\sigma^3, \sigma^0\}$ to the left of $\{\sigma^+, \sigma^-\}$ in the Weyl
$q$-spinor formulation and finally $\{\gamma^3,\gamma^0\}$ to the left of
$\{\gamma^1,\gamma^2\}$ in the Majorana $q$-spinor basis. Here $(\sigma^0,
\sigma^3, \sigma^+, \sigma^-)$ and $(\gamma^0, \gamma^1, \gamma^2, \gamma^3)$
are $q$-deformed Pauli and Dirac matrices \cite{Ca2}. Additionally $\tilde{f}$
is the Fourier transform of f', i.e.\

{\large\begin{equation}
\tilde{f}(\lambda_i,\lambda_j)=(2\pi)^{-2}\int^\infty_{-\infty}
d\mu_id\mu_jf'(\mu_i \mu_j)e^{-i\mu_i\lambda_i}e^{-i\mu_j\lambda_j}
\end{equation}}
$$
i,j=(1,2),(3,4),(5,6).
$$
Then carrying on integration on $\lambda_1$, $\lambda_3$,
$\lambda_5$, $\lambda_6$ we obtain
{\large\begin{equation}
\int \int f=\int^\infty_{-\infty}d\lambda'_1d\lambda'_2d\lambda_2
\tilde{f}(0,\lambda_2)\delta(\lambda'_2(1-w^{-1})-\lambda_2e^
{-2i\lambda'_1 Q'})+
\end{equation}
$$
\int^\infty_{-\infty}d\lambda'_3d\lambda'_4d\lambda_4
\tilde{f}(0,\lambda_4)\delta(\lambda'_4(1-w^{-1})-\lambda_4e^
{-2i\lambda'_3 Q'})+
$$
$$
\int^\infty_{-\infty}
d\lambda'_5d\lambda'_6\tilde{f}(0,0),
$$ }
that after changing the order of integration and integrating on
$\lambda_2$ and $\lambda_4$ becomes
{\large\begin{equation}
\int \int f=\int^\infty_{-\infty}d\lambda'_1d\lambda'_2e^{\lambda'_1 Q'}
\tilde{f}(0,\lambda'_2(1-w^{-1})e^{2i\lambda'_1 Q'})+
\end{equation}
$$
\int^\infty_{-\infty}d\lambda'_3d\lambda'_4e^{-\lambda'_3 Q'}
\tilde{f}(0,\lambda'_4(1-w^{-1})e^{-2i\lambda'_3 Q'})+
$$
$$
\int^\infty_{-\infty}d\lambda'_5d\lambda'_6\tilde{f}(0,0)
$$}
The last term in eq(36) corresponds to the ordinary divergent term
that appears in the standard commutative algebraic formulation of quantum
field theory; there is no way we can recover a finite term out of this
in the limit $q\rightarrow 1$. Checking the non-commutative Hopf algebra
generated by $X^{ij}$ we find the reason why this happens to be so; T is
central with respect to $(X,Y,Z)$, so this part of the Haar measure is not
really defined on a non-commutative algebraic variety. Therefore we can
extract out of $\int\int f$ a $q$-regularizable part
\begin{equation}
\int \int f-\int^\infty_{-\infty}d\lambda'_5d\lambda'_6\tilde{f}(0,0)=
\end{equation}
$$
(2\pi \delta (0)) \left\{\int^\infty_{-\infty}d\lambda'_2
\tilde{f}(0,\lambda'_2(1-w^{-1}))+
\int^\infty_{-\infty}d\lambda'_4
\tilde{f}(0,\lambda'_4(1-w^{-1}) \right\}.
$$
But $lim_{q\rightarrow 1}w^{-1}=1$ thus, as $q\rightarrow 1$,
$\int \int f-\int^\infty_{-\infty}d\lambda'_5d\lambda'_6\tilde{f}(0,0)$
diverges, by contrast at $q\neq 1$ and assuming suitable analicity and decay
of $\tilde{f}$ to allow contour integration, eq(37) can be made finite
for suitable f; moreover, this is proportional to $(1-w^{-1})^{-1}$.\

In the limit $q\rightarrow 1$, the transformation described in eq(26) is non
sense, because in this limit the map ${\it H}\rightarrow {\it L}$ is
singular. We remark that this does not mean that the $q$-regularization
scheme performed on the algebra ${\it H}$ and described up to here is lacking
sense in case $q=1$ but only the map that relates this with
$A^{2/0}_q\otimes A^{2/0}_q$. We are interested in this map because might be
of some help in the future construction of $q$-regularization on $q$-Minkowski
space-time. Further work should be done in this direction.\

Q.E.D.\

For obvious reasons, the vertex correction for $\lambda\phi^4$ theory described
in example 1 is suitable of being $q$-regularized on ${\it H}$ as well as in
example 2 was on ${\it L}$. Further work should be done to generalize these
examples to more interesting cases. Since this scheme is strongly basis
dependant,
a complete analysis of the class of functions suitable of being $q$-regularized
on physically interesting basis is needed. Note that $q$-regularization may
be considered equivalent to dimensional regularization in a similar sense to
the McKane and Parisi-Sourlas case \cite{MPS}.\

\section{Comments and Remarks.}

In the paper wherein Woronowicz \cite{Wo} proves the existence and uniqueness
of the Haar measure, i.e. the unique state invariant under left
(and simultaneusly right) shifts, for any compact quantum group, he proposes
the following $q$-integration on $SU_q(2)$
\begin{equation}
\int^1_{q^0}{\bf f}=(1-q)\sum^\infty_{k=0} q^k{\bf f}(q^k)
\;\;\;\mbox{for any  } {\bf f}\in SU_q(2),
\end{equation}

On the other hand, let us set $T=0$ in eq(25) and from eq(7), we get
\begin{equation}
U_2(\hbar {\bf u},0)=q^{\frac{1}{2}}\tilde{u}_1(0,\hbar {\bf u})
\;\mbox{ and }\;U_2(0,\hbar {\bf u})=-q^{-\frac{1}{2}}
\tilde{u}_1(\hbar {\bf u},0),
\end{equation}
this is equivalent to set ${\tilde Z}^{\rho}$$=\epsilon^{\rho\sigma}
{\bar Z}_{\sigma}$ in eq(9).
Thus
$$
A=X+Y,\;\;\;\bar{A}=X-Y\;\;\;B=\bar{B}=Z.
$$

We call $X^1=\frac{1}{2}(A+\bar{A})$,$X^2=\frac{1}{2}(A-\bar{A})$ and
$X^3=B$; then in terms of this $(X^1, X^2,X^3)$ 3-dimensional vector
representation we propose the following basis to be used;
\begin{equation}
F^{\lambda_1 \lambda_2 ,\lambda_3 \lambda_4}=
\left(
e^{i\lambda_1x^3}e^{i\lambda_2x^1},e^{i\lambda_3x^3}e^{i\lambda_4x^2}
\right)
\end{equation}
where $X^1=x^1$, $X^2=x^2$, and $X^3=q^{\frac{x^3}{Q'}}$, as was done  in
eq(26).\\

{}From the work done on the category of representations of a Hopf algebra we
can write the action of any function f of the Hopf algebra $SU_q(2)$ on its
vector representation V, through the corresponding basis;
\begin{equation}
{\bf f}.e^{j}_{m}=\sum_{i=+,-,0}f^{i}{\bf e_i}.e^{j}_{m}\in V
\;\;\;\;\;\;\forall {\bf f}\in SU_q(2)
\end{equation}
where ${\bf e_{+}}$=$X^{+}$,${\bf e_{-}}$=$X^{-}$, ${\bf e_{0}}$=$H$ is the
$SU_q(2)$ basis and $f^{i}\in C$.\

{}From eq(41) is clear that the Woronowicz's map $\int {\bf f} \rightarrow C$
for the $SU_q(2)$ Haar measure induces a $\int \int f\rightarrow C$ map for
the vector representation of the Hopf algebra $SU_{q}(2)$, inducing another
for $\Lambda $ which is written in terms of $SU_q(2)$. We define the
$\Lambda$ matrix as
\begin{equation}
\Lambda^{(ij)}_{(i'j')} \equiv \tilde{M}^{i}_{i'}M^{j}_{j'}
\;\;\;\;;\;\;M\in SL_q(2,C),\; \tilde{M}\in \tilde{SL}_q(2,C)\;.
\end{equation}
Thus, the similarity of eq(34) and eq(38) can be understood in these terms.\

{}From these two facts and the obvious similarity of the $q$-integration
carried out in eq(34) and the one depicted in eq(38), we think that the
$q$-Time zero-projection in the $q$-deformed space defined for the second
example corresponding to the $\tilde{M}^{\dagger}=M^{-1}$ identification,
reduces $\int\int f-\int^{\infty}_{-\infty}d\lambda'_{5}d\lambda'_{6}
\tilde{f}(0,0)$ in Theorem 2 to the Haar weight on the vector representation
of $\Lambda$ written in terms of $SU_q(2)$.\

Finally, we show how null directions in $\Lambda$ can lead us to obtain the
quantum mechanical Galilei group. By imposing the following null bi-ideals
$$
u^1_2=0,\;\;\;u^2_1=0,
$$
$$
u^1_1u^2_2=u^2_2u^1_1=1,
$$
being $u^i_j\in M$ ( the same is asked for ${\tilde u}^i_j$ being
${\tilde u}^i_j\in {\tilde M}$), we obtain a direct product representation of
the quantum Galilei group.\

We can see this from the viewpoint of cohomological formalism. We construct
the quantum mechanical Galilei group, choosing the following Galilei
transformations
\begin{equation}
{\bf r'}={\bf r}+{\bf v}t\;\;\;\;\;{\bf p'}={\bf p}+m{\bf v}
\end{equation}
where m is the particle mass.\\

Then eq(2) transforms to
{\large\begin{equation}
U({\bf v})=e^{i\frac{{\bf v}\cdot ({\bf p}t-m{\bf r})}{\hbar}},
\end{equation}}
and its action on a wave function $\Psi({\bf r})$ introduces a phase
(one-cocycle) $\alpha_1$, i.e.
\begin{equation}
U({\bf v})\cdot \Psi({\bf r})=e^{2i\pi \alpha_1({\bf r};{\bf v})}
\cdot\Psi({\bf r}+{\bf v}t).
\end{equation}
We shall consider this one-cocycle as trivial, so
\begin{equation}
\alpha_1(\bf {r;v})=\delta \alpha_0=\alpha_0({\bf r'})-\alpha_0({\bf r})
\end{equation}
where $\alpha_0$ is a function, called 0-cocycle, which depends only on
{\bf r}.\

Therefore, the group law of the quantum mechanical Galilei group
for translations on phase space (or $U(1)$ extended Galilei group) is
expressed such that
\begin{equation}
e^{2i\pi \alpha_0({\bf r'})}=e^{2i\pi (\alpha_0({\bf r})+\phi+\alpha_1({\bf
r;v}))}
\end{equation}
where
\begin{equation}
2\pi \alpha_1=\frac{1}{\hbar}(m{\bf v}\cdot {\bf r}+\frac{1}{2}mv^2t)
\end{equation}
and $\phi$ is the central parameter of the quantum mechanical Galilei group.\

On the other hand, we require
\newline
$M=\left(\begin{array}{cc}
u^1_1 & u^1_2 \\
u^2_1 & u^2_2
\end{array}
\right)\in SL_q(2) \mbox{ and } $
$M=\left(\begin{array}{cc}
{\tilde u}^1_1 & {\tilde u}^1_2 \\
{\tilde u}^2_1 & {\tilde u}^2_2
\end{array}
\right)\in {\tilde SL}_q(2)$ to belong to the quantum mechanical Galilei
group; i.e. $M$ ( equivalently ${\tilde M}$) must fulfill eq(5). It is
straightforward to prove that, in this case, the following null bi-ideals
have to be imposed on $M$ (equivalently on ${\tilde M}$)
\begin{eqnarray}
u^1_2=0 & & u^2_1 = 0 \\
u^1_1u^2_2 & = & u^2_2u^1_1=1 ,\nonumber
\end{eqnarray}
so to end up with a group that has only one generator as should be.\

Besides, we can prove that the null bi-ideals once imposed on $M\in SL_q(2)$
(thereby defining the quantum mechanical Galilei group) produce the following
pairing
\begin{eqnarray}
<u^1_2,t^{\dagger  k}_m>=R^{1k}_{2m}=0 \;\;\;k,m,s=1,2 \nonumber \\
<u^2_1,t^{\dagger  k}_m>=R^{2k}_{1m}=0 \\
<u^1_1u^2_2,t^{\dagger  k}_s>=R^{1k}_{1m}R^{2m}_{2s}=1 \nonumber \\
<u^2_2u^1_1,t^{\dagger  k}_s>=R^{2k}_{2m}R^{1m}_{1s}=1 \nonumber
\end{eqnarray}
where $t^{\dagger  k}_m \;\; (k,m=1,2)$ are generators of the dual Hopf
algebra for $SL_q(2)$ and $R^{ij}_{kl} \;\;(i,j,k,l=1,2)$ entries of
 the Yang-Baxter matrix $R_{G}$ associated with the quantum mechanical
Galilei group. This does not determine $R_{G}$ but restricts the solution
to block diagonal matrices.\

Finally, if we impose the null-directions given by eq(50) in $\Lambda$ we
obtain the following representation of the quantum Galilei group \newline

{\large\begin {equation}
\Lambda=
\left(
\begin{array}{cccc}
(\tilde{u}^1_1)^{-1}u^1_1 & 0 & 0 & 0 \\
0 & \frac{\tilde{u}^1_1u^1_1+q^2(\tilde{u}^1_1u^1_1)^{-1}}{1+q^2} & 0 &
\frac {q^2(\tilde{u}^1_1u^1_1-(\tilde{u}^1_1u^1_1)^{-1})}{1+q^2} \\
0 & 0 & \tilde{u}^1_1(u^1_1)^{-1} & 0 \\
0 & \frac{\tilde{u}^1_1u^1_1-(\tilde{u}^1_1u^1_1)^{-1}}{1+q^2} & 0 &
\frac {q^2(\tilde{u}^1_1u^1_1+(\tilde{u}^1_1u^1_1)^{-1})}{1+q^2}
\end{array}
\right)
\end{equation}}

Summarizing, in this paper we have introduced the concept of $q$-regularization
and used the projective representation of the non-commutative Heisenberg
algebra to construct the Manin quantum plane, thereby defining $q$-spinors.
Using this as a building block we present a $q$-regularization in terms of a
four dimensional representation of a particular two dimensional non-commutative
space
at first. Besides, regularization on a $q$-deformed space that can be maped
into a particular braided product of Manin's quantum plane and it is related
to the first $q$-space is studied too. We show how to extract, from relevant
quantities, finite components (provided $q\neq 1$) that can become infinite
at $q=1$. To compute the Haar weight, we propose particular basis projected
from $q$-deformed spaces, so the functions to be $q$-regularized are to be
considered on this frame of reference. An example for $\lambda\phi^4$ field
theory is presented. Additional work must be done to generalize our scheme
to any arbitrary function on $q$-Minkowski space-time basis.\

Finally, in order to learn about the general scheme and its symmetries, we
study the T=0 Haar measure in terms of the $SU_q(2)$ measure and the null
directions in the Hopf algebra that lead to a quantum mechanical Galilei
group.\

Although in this paper we can $q$-regularize only a class of suitable
functions (restricted by the particular basis chosen), we think that the full
prescription, derived from physical considerations, might apply to make
relevant quantities in field theory finite at $q\neq 1$.

\section{Acknowledgments}

This work was partially supported by CONACYT, M\'exico. I wish to thank ITD
in UC Davis for its hospitality

\end{document}